\documentclass[runningheads]{llncs}
\usepackage[T1]{fontenc}
\usepackage{graphicx,verbatim}
\usepackage{amsmath}
\usepackage{multirow}
\usepackage{booktabs}
\usepackage{cite}
\usepackage{tabularx}
\usepackage{makecell}
\begin{document}
\title{3D Heart Reconstruction from Sparse Pose-agnostic 2D Echocardiographic Slices}
%

\author{%
  Zhurong Chen\inst{1} \and
  Jinhua Chen\inst{1} \and
  Wei Zhuo\inst{1} \and
  Wufeng Xue\inst{1}\thanks{Corresponding author} \and
  Dong Ni\inst{1}}

\authorrunning{Chen et al.}  

\institute{%
  Shenzhen University, Shenzhen, China\\
  \email{xuewf@szu.edu.cn}}


\maketitle              
\begin{abstract}
Echocardiography (echo) plays an indispensable role in the clinical practice of heart diseases. However, ultrasound imaging typically provides only two-dimensional (2D) cross-sectional images from a few specific views, making it challenging to interpret and inaccurate for estimation of clinical parameters like the volume of left ventricle (LV). 3D ultrasound imaging provides an alternative for 3D quantification, but is still limited by the low spatial and temporal resolution and the highly demanding manual delineation. 
 
To address these challenges, we propose an innovative framework for reconstructing personalized 3D heart anatomy from 2D echo slices that are frequently used in clinical practice. Specifically, a novel 3D reconstruction pipeline is designed, which alternatively optimizes between the 3D pose estimation of these 2D slices and the 3D integration of these slices using an implicit neural network, progressively transforming a prior 3D heart shape into a personalized 3D heart model. 

We validate the method with two datasets. When six planes are used, the reconstructed 3D heart can lead to a significant improvement for LV volume estimation over the bi-plane method (error in percent: 1.98\% VS. 20.24\%). In addition, the whole reconstruction framework makes even an important breakthrough that can estimate RV volume from 2D echo slices (with an error of 5.75\% ). This study provides a new way for personalized 3D structure and function analysis from cardiac ultrasound and is of great potential in clinical practice.



\end{abstract}
\section{Introduction}

Cardiovascular diseases remain the leading cause of death worldwide, emphasizing the critical importance of accurate cardiac function assessment. Using traditional two-dimensional (2D) ultrasound imaging, is challenging to interpret and inaccurate for 3D heart structure and function analysis. For example, Simpson's biplane method is most widely used for the calculation of LV volume. It assumes that the cavity of LV is composed of a series of eclipse-shaped disks~\cite{kircher1991left}, which may not be accurate and lead to underestimation of the LV volume. The situation is even worse for right ventricle (RV) volume, who has a complex shape that the volume cannot be well estimated only from 2D planes. 
3D ultrasound provides an alternative for 3D cardiac function analysis, but is still limited by the low spatial and temporal resolution. Besides, 3D quantification analysis requires heavy manual adjustment and takes much time even with the help of semi-automatic software. 

3D reconstrcution from multiple 2D images has been well explored for nature scene and provides a solution for our problem. 
In the medical area, statistical shape models (SSM) and deep learning-based methods have achieved remarkable progress in 3D heart modeling~\cite{zhang2020review,freitas2024automatic,stojanovski2022efficient,laumer2023fully}. SSM-based approaches~\cite{freitas2024automatic} leverage statistical constraints to reconstruct heart structures from sparse data, but heavily rely on high-quality datasets and is limited to capture individual anatomical variations. End-to-end methods based on deep learning, such as those utilizing convolution neural networks~\cite{stojanovski2022efficient,laumer2023fully}, graph neural networks~\cite{laumer2023weakly} or implicit neural representation (INR)~\cite{ma2023cardiacfield}, can directly generate 3D heart models from multiview 2D slices. However, these methods are sensitive to data sparsity and misaligned slice positions, leading to suboptimal performance under challenging conditions. 
Moreover, existing methods usually rely on well-aligned or position-aware 2D slices, whereas the accurate slice position in the 3D space is usually not known in practice. 

To address these challenges, we propose a novel 3D heart reconstruction method (Echo3D), which alternatively iterates between estimation of the 3D position of the input 2D slices and the reconstruction of the personalized 3D heart shape using implicit neural representation (INR)-based representations.
In this work, the required multiple slices of 2D heart structures can be easily obtained by applying the state-of-the-art segmentation model to the ehco planes. Starting from these segmented slices whose 3D positions are not accurately known, our method can not only achieve anatomically consistent personalized 3D heart reconstruction, but also enables precise estimation of  LV and RV volumes. In addition, the method can also be employed for 3D cardiac structure and function analysis such as regional wall thickness estimation, and strain analysis.  

The main contributions of this study are as follows.
\begin{itemize}
    \item  We introduced a new paradigm for evaluation of 3D heart structure and function from multiple 2D echo slices.
    \item We proposed a novel ultrasound-based 3D heart reconstruction method, Echo3D, where multiple position-agnoistic 2D echo planes can be employed to accurately reconstruct the 3D heart shape.

    \item We conducted extensive evaluations in terms of 3D shape reconstruction, and volume estimation of LV and RV. The results show its superior performance over the widely-used biplane method and existing methods.
\end{itemize}

\section{Related Work}

\subsection{3D Pose Estimation of 2D slices}
Pose estimation, which involves determining the spatial configuration of 2D slices within a 3D coordinate system, is a crucial step in 3D reconstruction. Existing approaches have explored various strategies across ultrasound and MRI modalities to address this challenge. Freitas et al. \cite{freitas2024automatic} proposed an automatic multiview pose estimation framework for focused cardiac ultrasound, leveraging motion tracking and machine learning techniques to align multiple 2D views in 3D space. Yang et al. \cite{yang2021searching} introduced a collaborative multi-agent system for anatomical plane localization in 3D ultrasound. Later, Yang et al. \cite{yang2021agent} enhanced this method with adaptive dynamic termination strategies, improving efficiency and precision for clinical applications. Despite their good performance, these methods require either extra sensors or the 3D volume as a reference, and are not applicable to our case.

\subsection{3D Reconstruction from Pose-Aware 2D slices} 
Traditional statistical shape models (SSM)~\cite{unberath2015open}, relied on prior anatomical knowledge for reconstruction, but struggled to capture individual anatomical variability. Furthermore, SSM approaches required extensive labeled datasets, limiting their use in personalized modeling.

Deep learning methods have significantly advanced 3D reconstruction accuracy. 
Ma et al. \cite{ma2023cardiacfield} introduced CardiacField that can reconstruct the volume of 3D ultrasound from a large number of densely distributed 2D videos. \cite{sawdayee2023orex} introduced an INR-based approach to reconstruct 3D objects from dense planar cross sections. 
Despite their strong interpolation capabilities, they required dense 2D slices with accurate position information. 
Laumer et al. \cite{laumer2023weakly} developed a generative approach to infer personalized heart meshes from apical four chamber echo (A4C) videos. However, only the A4C view is used and the generative procedure may introduce incorrect structure into the final 3D shape. Beyond ultrasound, \cite{xu2023nesvor} introduced INR for slice-to-volume reconstruction in MRI, where the 2D slices are well aligned. 

\subsection{3D Reconstruction from Pose-agnostic 2D slices}
In the domain of free-hand ultrasound, INR-bsed methods have been employed for sensorless 3D ultrasound volume reconstruction from a large number of adjacent 2D slices~\cite{yeung2021implicitvol}, novel views synthesizes from multiple ultrasound planes~\cite{wysocki2024ultra}, and distance field-based 3D structure reconstruction from irregularly spaced slices~\cite{chen2024rocosdf}. 
However, the structure reconstruction relies on dense calibration and slice coverage, and uniform slice alignment. It faces challenges such as being computationally intensive and the requirement of uniform slice alignment.

\section{Methodology}

\subsection{New Paradigm for 3D Cardiac Analysis}
\begin{figure}[tbp]
    \centering
    \includegraphics[width=0.85\textwidth]{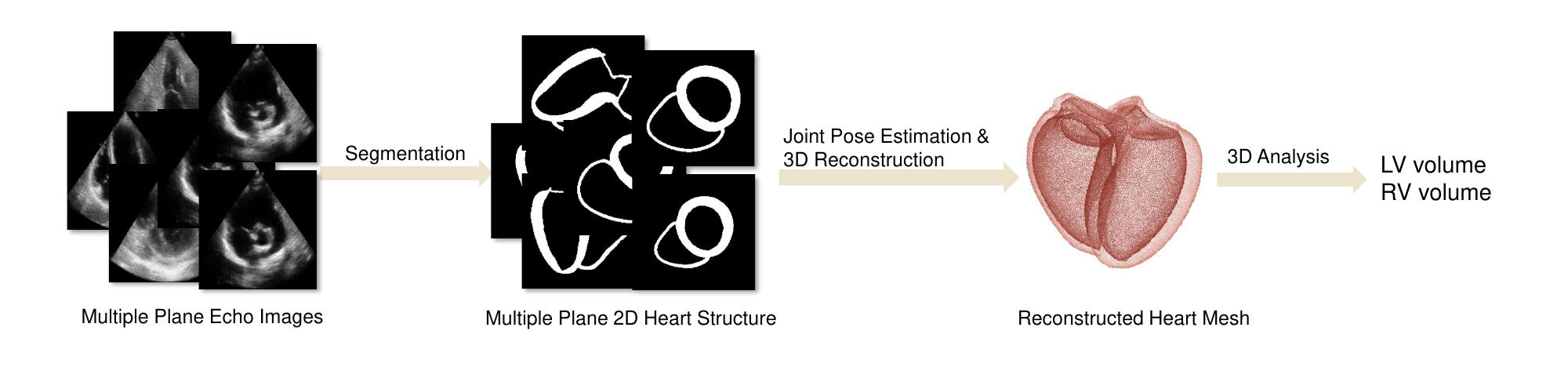} 
    \caption{The new paradigm consists of four stages: 1) acquisition of multiple 2D echo images, 2) segmentation of heart structures from these planes, 3) joint pose estimation and 3D heart reconstruction, and 4) 3D analysis for LV and RV volumes.} 
    \label{fig_paradigm_3stage} 
\end{figure}

We present a new paradigm for 3D cardiac structure and function analysis from ultrasound imaging (Fig.\ref{fig_paradigm_3stage}), which includes 1) imaging of multiple 2D echo planes, 2) segmentation of these 2D planes, 3) 3D heart reconstruction from the segmented results, and 4) 3D quantification and analysis.  

For 2D echo planes, instead of using a large set of echo planes recommended by~\cite{mitchell2019guidelines}, we consider only the ones most commonly used in clinical practice: apical (AP) views of 2 chamber (2CH), 3 chamber (3CH) and 4 chamber (4CH), and parasternal short-axis (PSAX) views of mitral valve, papillary muscle and apical levels. This simplification enables the method to be easily deployed for real-world applications. Segmentation of echo images has been well investigated~\cite{liu2021deep, painchaud2022echocardiography, chiou2021ai,wei2023co}, and will not be the focus of this work. In our work, we first utilize the 3D heart mesh cohort~\cite{rodero2021linking} to generate 2D heart structures from views that are consistent with the six planes mentioned above. 
Then we dedicate ourselves to the 3D heart reconstruction in the following.

\subsection{3D Heart Shape Reconstruction}

\begin{figure}[tbp]
    \centering
    \includegraphics[width=0.95\textwidth]{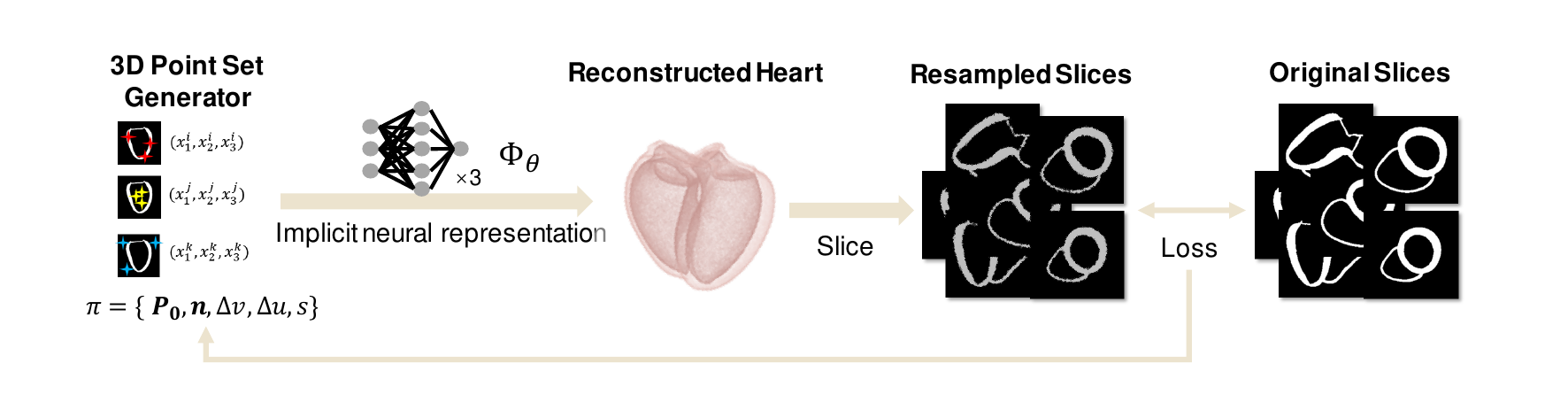} 
    \caption{3D heart reconstruction from position-agonistic 2D slices. } 
    \label{fig_Pipeline} 
\end{figure}

The main challenge for the reconstruction in our case is that these 2D structure slices are sparsely distributed in the 3D space and their pose in the 3D space are not accurately aware. 
Consider a set of $N$ 2D structure slices $\{S_1, S_2, \dots, S_N\}$ with unknown spatial pose, the objective is twofold: (1) estimate the spatial pose $\pi_j$ of each slice $S_j$, and (2) reconstruct a complete and anatomically consistent 3D heart structure $H_{rec}$ that aligns with these 2D slices when well-posed. Denote the the reconstruction model as $\Phi_\theta$, where $\theta$ is the parameters to be optimized, the problem can be defined as: 
\begin{align}
    \text{minimize}_{\theta, \{\pi_j\}} \quad
    & \sum_{j=1}^{N} \mathcal{L}(Slice(H_{rec},\pi_j), S_j)
    \label{eq_optimize}
\end{align}
where $H_{rec}$ is obtained by applying $\Phi_{\theta}( )$ to all locations of the 3D grid, $\mathcal{L}$ calculates the loss between the points of two structure slices, and $Slice(\cdot, \pi)$ extracts the cross-section of a 3D structure with pose specified by $\pi$. 

This optimization can be achieved through an alternative optimization iteration between $\theta$ and $\{\pi_1\,\cdots,\pi_N\}$, with a well-designed initial state according to prior knowledges of heart shape. 
Updating $\theta$ in each iteration can be viewed as deforming the prior heart shape to fit the spatially positioned 2D slices. This design combines the benefits of a global shape prior with local refinements, ensuring both anatomical consistency and accurate surface modeling.

\subsection{3D Reconstruction by Optimizing $\Phi_{\theta}$}

In our work, the MLP-based implicit neural representation is used to represent the 3D heart structure. Specifically, the implicit function $\Phi_{\theta}(x_i)$ outputs for a 3D coordinate $x_i$ a scalar value $c_i$ that is constrained to $[-1,1]$. \(c < 0\) indicates points inside the shape, \(c = 0\) identifies points on the surface, and \(c > 0\) corresponds to points outside the shape.

With the previously estimated slice poses $\pi_j^{k-1}$, the reconstruction can be solved well by matching the output of the implicit function to the value of the points that lie in the input 2D slices. 
\begin{align}
    \theta^{k} = \text{minimize}_{\theta} \sum_i \mathcal{L}(\Phi_{\theta}(x_i^{k-1}), z_i) 
\end{align}
where $x_i^{k-1}$ is a point from any of the input slices that is transformed into the 3D space with $\pi^{k-1}$, and $z_i$ is the value similar to $c$ for points on these slices. 

\begin{figure}[tbp]
    \centering
    \includegraphics[width=0.95\textwidth]{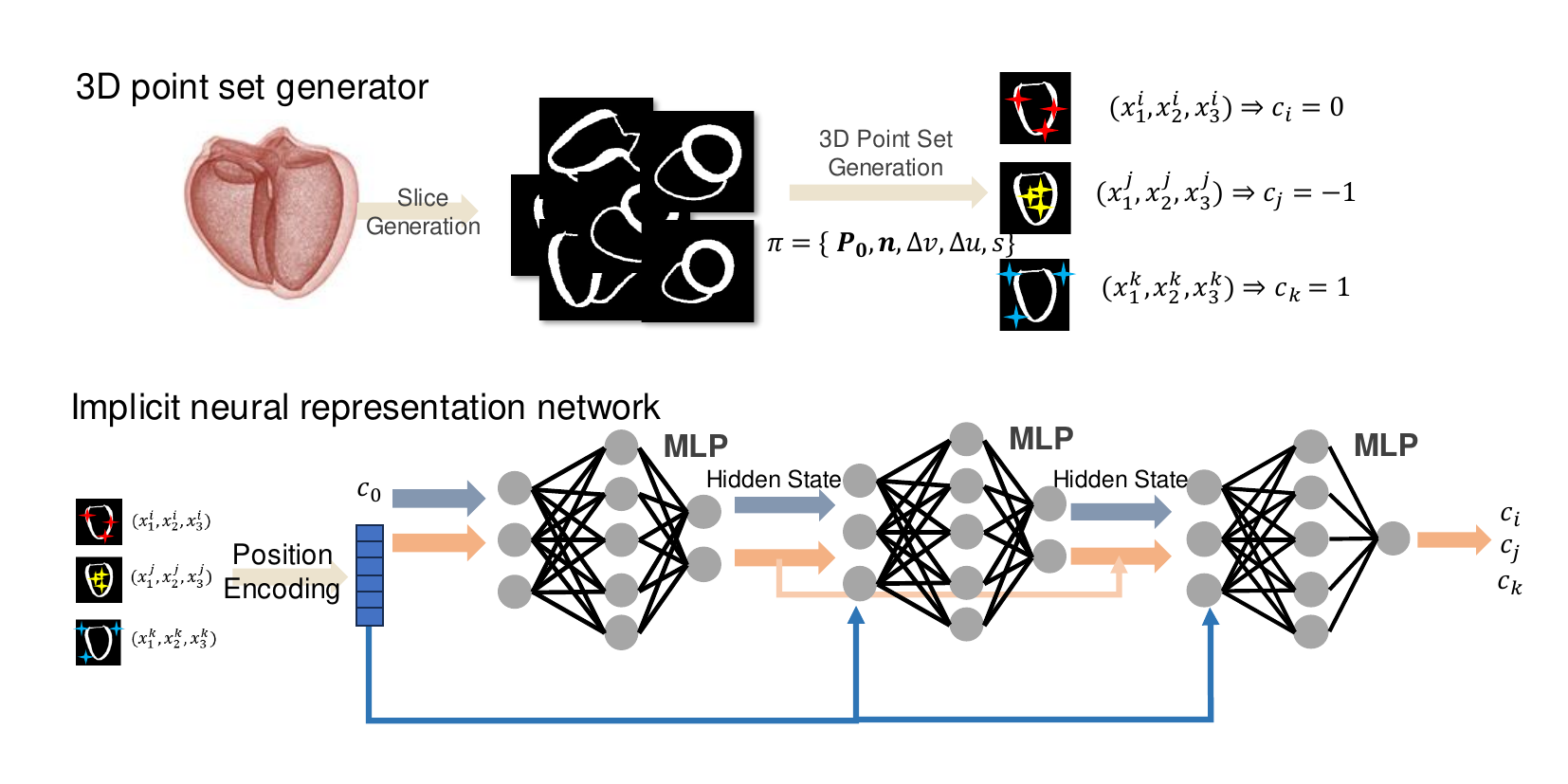} 
    \caption{The point set generator (top) and the implicit neural representation network (bottom) used in our reconstruction. } 
    \label{fig_MethodDetail} 
\end{figure}

During optimization, a point set generator (top of Fig.\ref{fig_MethodDetail}) is designed to generate 3D point sets $\{z_i\}$ uniformly from the 2D slices, and to map these 2D slices into 3D coordinates. 
 
\subsection{Pose Estimation by Optimizing $\pi$}
The pose to be optimized for each slice contains a set of parameters: $\pi=\{\mathbf{P_0}, \mathbf{n},\Delta v, \Delta u, s\}$, where $\mathbf{P_0}$ is the reference point, $\mathbf{n}$ denotes the normal vector, $\Delta v, \Delta u, s$ represents the translation and scaling factor.
For pose estimation of slice $S_j$ given the previous reconstructed model $\Phi_{\theta^{k-1}}$, the objective is: 
\begin{align}
    \{\pi_1,\cdots,\pi_N\}^k=\text{minimize}_{\{\pi_1,\cdots,\pi_N\}} \sum_{j=1}^N\sum_{x_i\in S_j} \mathcal{L}(\Phi_{\theta^{k-1}}(x_i), z_i) 
\end{align}
This optimization involves adjusting the orientation with the normal vector, scaling the plane via \(s\), and translating its position using \(\Delta u\) and \(\Delta v\).

\subsection{Implementation Details}
In this work, the implicit function is implemented as MLP, and the loss function in Eq.~\ref{eq_optimize} combines a margin-based reconstruction error term $\mathcal{L}_{\text{Rec}}$ and a smoothness regularization $\mathcal{L}_{\text{SM}}$:
\begin{align}
\mathcal{L}_{\text{Rec}}(x_i,c_i) = \max(0, 1 - c_i \cdot (\Phi_{\theta}(\mathbf{x}_i)), 
\end{align}
\begin{align}
[\mathcal{L}_{\text{SM}}\ =\ \lambda\ \max(0,\ |\nabla\ \Phi(\mathbf{x})|\ -\ \alpha)],
\end{align}
where $c_i$ is the ground truth label, \(\nabla \Phi(\mathbf{x})\) denotes the gradient of the scalar field, \(\lambda\) controls the regularization strength, and \(\alpha\) defines a threshold for the gradient magnitude.
Inspired by the work of \cite{mildenhall2021nerf} and \cite{sitzmann2020implicit}, positional encoding is used for the input of the implicit function $\Phi_{\theta}$.

\section{Experiment}

\subsection{Dataset}
In this study, we utilize two datasets of 3D heart models with four chambers to generate the six sparse 2D slices and to evaluate the reconstruction performance. 

\subsubsection{Real Heart Meshes from CT Images}
The first database of healthy four-chamber heart models~\cite{rodero2021linking} is explicitly designed for electromechanical (EM) simulations. It includes 20 four-chamber heart models derived from diastolic phase CT images of patients with acute chest pain in the emergency department. 

\subsubsection{Virtual Synthetic Heart Meshes}
This dataset comprises 1000 synthetic four-chamber heart models that are generated using a Statistical Shape Model (SSM) from the above 20 meshes. The variation includes differences in chamber volumes, wall thicknesses.
We randomly selected 40 models from this dataset.

\subsection{Evaluation Metrics}
The performance of the proposed method is evaluated in terms of 1) overlapping between volumes (IOU), 2) Chamfer distances between two surface~\cite{fan2017point}, and 3) estimation error of clinical parameters.

\begin{table}[b]
\centering
\begin{tabular}{ll|cccc|cc}
\hline
 \textbf{Method} & &\multicolumn{4}{c|}{\textbf{Echo3D}}  & \textbf{ORex} \\
 \hline
 \textbf{Input}   &    & A2C\&A4C & 3AP & A4C+3SAX & ALL6 & ALL6\\ 
 \hline
 \multirow{3}{*}{\textbf{Virtual}} 
& EP & 91.20  /  3.09         & 92.82  /  2.52        & 94.20 / 2.13       & 95.35  / 1.76     &92.85/2.42         \\  
& LV & 90.95 / 1.99         & 92.29 / 1.67        & 93.61 / 1.50           & 95.24  / 1.12     &93.80/1.38      \\  
& RV & 82.92 / 3.68         & 86.02 / 2.98        & 90.72 / 1.97           & 91.32  / 1.85     &87.10/ 2.59      \\ \hline
\multirow{3}{*}{\textbf{Real}} 
& EP & 83.75 / 5.30       & 88.25  / 3.80        & 86.93  / 4.22       & 90.42  / 3.05      &85.54/ 3.64 \\
& LV & 81.98 / 3.69        & 86.49  / 2.72        & 84.34  / 3.21       & 89.48  / 2.13     &88.37/ 2.61 \\
& RV & 73.63 / 5.24        & 78.86  / 4.11        & 79.46  / 4.00       & 84.11  / 3.04     &77.29/4.71   \\ \hline
\end{tabular}
\caption{Performance of the 3D reconstruction in terms of volume overlap and chamfer distance for LV, RV, and Epicardium (EP) when different input slices are used. For each cell, IOU (\%) and $d_{\text{CF}}$ (mm) are displayed.}
\label{tab_iou}
\end{table}

\section{Result and Analysis}


\subsection{Performance of 3D reconstruction}


Table~\ref{tab_iou} shows the reconstruction accuracy of Echo3D with different combination of 2D slices. With all six slices (ALL6), it achieves the best results for all regions, exceeding the IOU of 95\% for LV and EP on the virtual dataset. Despite the complex structure of RV, Echo3D can still achieve accurate reconstruction, with IOU of 91.32\% and 84.11\% for both datasets. When only A4C and A2C are used, the LV can still be well reconstructed, which implies its potential to estimate the 3D volume of the LV.
The last column of Table~\ref{tab_iou} shows the performance of OReX \cite{sawdayee2023orex} with six planes. It achieves a lower IOU and a higher distance than Echo3D does, despite the fact that it requires the position of the 2D slices. No effective 3D shapes can be obtained when the end-to-end data-driven method E-Pix2Vox++\cite{stojanovski2022efficient} is applied, which may be attributed to its demand of large datasets for training. Visualization results can be found in the Supplementary.


\begin{table}[t]
\centering
\begin{tabular}{llp{4cm}p{3cm}}
\hline
\textbf{Dataset}&\textbf{Method}       & \textbf{Volume Error (ml) \newline (LV/RV)} & \textbf{Relative error (\%) \newline (LV/RV)} \\ \hline
\multirow{5}{*}{\textbf{Virtual}}
&Simpson's Biplane & 17.92  /-           & 14.18 /-            \\ 
&OReX(ALL6)             & 5.58/14.66          &4.43/9.95       \\
\cline{2-4}
&Echo3D(A2C\&A4C)  & 6.24   /  10.48         & 4.94   / 9.46       \\ 
&Echo3D(3AP)            & 6.00 /  7.02           & 4.83 / 6.34        \\ 
&Echo3D(A4C+3SAX)           & 4.86 / 3.59        & 3.85 / 3.24       \\ 
&Echo3D (ALL6)          & 2.60 / 1.92         & 2.06 / 1.73        \\  \hline
\multirow{5}{*}{\textbf{Real}}
&Simpson's Biplane   & 25.03 / -  & 20.24 /-  \\ 
&OReX(ALL6)             & 6.58/31.01          &5.39/21.33 \\
\cline{2-4}
&Echo3D(A2C\&A4C)    & 9.85 / 10.52      & 8.09   / 9.64      \\ 
&Echo3D(3AP)        & 4.59 / 10.50      & 3.71   /  9.49           \\ 
&Echo3D(A4C+3SAX)   & 6.43 / 10.24     & 5.23    / 9.34         \\ 
&Echo3D (ALL6)       & 2.45 / 6.43         & 1.98 / 5.73              \\ \hline
\end{tabular}
\caption{Estimation error of LV/RV Volume by Echo3D with different input slices and the Simpson's Biplane method.}
\label{tab_LV_volume_error}
\end{table}

\subsection{LV/RV 3D Volume Estimation Error}

Table~\ref{tab_LV_volume_error} presents the volume estimation errors for both LV and RV. As the number of planes used increases, a lower estimation error can be achieved. Echo3D achieves the lowest errors when using all six slices (ALL6), with estimation errors of 2.45 ml for LV and 6.43 ml for RV on the Real dataset. Similar trend can be observed for the Virtual dataset. 
Note that RV has much more complex structures than LV, and there is currently no effective method to estimate RV volume from 2D echo images.
Another inspiring result is that for both datasets, Echo3D obviously outperforms the frequently used Simpson's biplane method, even when only A4C and A2C are used in our method. 
Visualization can be found in the Supplementary.


\subsection{3D reconstruction from real ultrasound planes.}
We apply Echo3D to segmentation results of real echo images from three subjects. The visualized results can be found in the supplementary. By Observing from different viewpoints, we can see that our model can well reconstruct the 3D heart model, variations in the global size, the apical shape, and detailed surface curve can be well captured.   

\section{Conclusion}
In this work, we presented a new paradigm for 3D analysis of heart structure and function from 2D echocadiography planes to take advantage of its high spatial and temporal resolution. To achieve this, we proposed an alternative optimization iteration method between pose estimation of 2D slices and 3D reconstruction. Experiments demonstrated that the method can accurately reconstruct LV, RV and EP with up to six commonly used planes, and shows great advantages in estimation of volume of LV and RV. 
The ability to reconstruct 3D heart structures enables comprehensive cardiac analysis, including wall thickness modeling, ventricular remodeling assessment, and 3D strain quantification. 



%
%
%
\bibliographystyle{splncs04}
\bibliography{ref}
%




\end{document}